\documentclass[a4paper]{article}

\usepackage[english]{babel}
\usepackage[utf8x]{inputenc}

\usepackage{booktabs}
\usepackage{tabu}
\usepackage[T1]{fontenc}
\usepackage{amssymb}
\usepackage[a4paper,top=3cm,bottom=2cm,left=3cm,right=3cm,marginparwidth=1.75cm]{geometry}
\usepackage{float}
\usepackage{amsmath}
\usepackage{graphicx}
\usepackage[colorinlistoftodos]{todonotes}
\usepackage[colorlinks=true, allcolors=blue]{hyperref}

\begin{document}
\begin{titlepage}
   \begin{center}
       \vspace*{1cm}


       \vspace{0.5cm}
       \LARGE
        \textbf{Analysis of ELO Rating Scheme in MOBA Games}
            
       \vspace{1.5cm}
        \normalsize
       Yuhan SONG\\
        \vspace{0.5cm}

       \vfill
            
            
       \vspace{0.8cm}

       \vspace{1cm}
        \Large
            
   \end{center}
\end{titlepage}


\setlength{\parindent}{0pt}

\section*{Summary of the Proposal}

ELO rating system is proposed by Arpad Elo, a Hungarian-American physics professor\cite{elo1978rating}. Originally, it was proposed for the ranking system of chess players, but it was soon adapted to many other zero-sum sports fields like football, baseball, basketball , etc. Nowadays, besides the traditional sports games, computer/video games are also playing an important role in social lives especially among the teenagers. In most of the online competition games, player's performance is usually scored and recorded by the game's ranking system. Meanwhile, ranking system like ladder in Dota is  not the only the metric for the players to evaluate their gaming strength, an ELO rating score based on players in-game performance is also a decisive factor for gamers' matching. Namely, the matching system will refer to players' score in the ranking system and performance score system to ensure the matched players will promisingly undergo a balanced game without one team dramatically overwhelming the other. ELO scheme and its variants in modern online competition games aims to ensuring the expected winning rate for each team approaches 50\%. However, ELO rating is also causing compliments among players. In this research, I will dig into the advantages and drawbacks of leveraging ELO ranking system in online games and why it is still employed by game developers despite the fact that it is disliked by most of the players. Also, a new effort based rating scheme will be proposed and compared with ELO scheme under the simulation environment.

\section*{Background}

Back to 1950s, the United States Chess Federation(USCF) was using a numerical rating system devised by Kenneth Harkness. This numerical system plays of role of tracking the individual wins and losses for each chess player. The basic principle of this ranking system is that when a player competes in a tournament, the average rating of his competition is calculated. If the player scores 50\% he receives the average competition rating as his performance rating. If he scores more than 50\% his new rating is the competition average plus 10 points for each percentage point above 50. If he scores less than 50\% his new rating is the competition average minus 10 points for each percentage point below 50\% \cite{enwiki:1050416173}. The original rating system was considered inaccurate under some circumstances. So in 1959, the USCF asked Elo to improve the Harkness rating system. Elo then proposed his new system in the same year, so that players’ ratings wouldn’t deviate much from the numbers they were used to. According to his new system an average player was rated 1500, a strong Chess club player 2000 and a grandmaster 2500. A key idea serves in ELO ranking system is that the difference in the ratings between two players serves as a predictor of the outcome of a match. Two players with equal ratings who play against each other are expected to score an equal number of wins. A player whose rating is 100 points greater than their opponent's is expected to score 64\%; if the difference is 200 points, then the expected score for the stronger player is 76\%. A player's Elo rating is represented by a number which may change depending on the outcome of rated games played. After every game, the winning player takes points from the losing one. The difference between the ratings of the winner and loser determines the total number of points gained or lost after a game. If the higher-rated player wins, then only a few rating points will be taken from the lower-rated player. However, if the lower-rated player scores an upset win, many rating points will be transferred. The lower-rated player will also gain a few points from the higher rated player in the event of a draw. This means that this rating system is self-correcting. Players whose ratings are too low or too high should, in the long run, do better or worse correspondingly than the rating system predicts and thus gain or lose rating points until the ratings reflect their true playing strength. The basic formulation for calculating the player's ranking score is described as below\cite{elo1978rating}:\\

\noindent Prior to the game, expected scores of the game opponents are calculated using the player's current ratings. Assuming player A's current rating is $R_A$, same for player B as $R_B$, the expected scores for A and B are:
    \[E_A=\frac{1}{1+10^{(R_B-R_A)/400}}\]
    \[E_B=\frac{1}{1+10^{(R_A-R_B)/400}}\]

These scores represents an expectation of winning for each player. For example, the expected score of 0.75 could represent a 75\% chance of winning and 25\% chance of losing. One thing worth noting is that there is not specific representation of drawing in the ELO rating system. The chance of drawing is embedded in the rate of win or loss. Taking the previous example into consideration, the expected score of 0.75 could also imply a 50\% chance of winning and 50\% chance of drawing.\\

Elo has introduced a adjustment factor $K$ for the players of different levels, for example $K=16$ for masters and $K=32$ for less competitive players.\\

Suppose player A scores $S_A$ points through a series of chess contest or through just single games with $K=32$, the updated ranking score of player A should become:
    \[R_A^{new}=R_A+K\cdot(S_A-E_A)\]

For example, if $R_A=2100$, $R_B=2000$ and $K=10$, and player B defeated player A in one game. Then the new ranking score of each player will be recalculated using the following parameters based on the former assumption: $E_A=0.64$, $E_B=0.36$, $S_A=0$, $S_B=1$, then: 
    \[R_A^{new}=R_A+K\cdot(S_A-E_A)=2100+10\cdot(0-0.64)=2094\]
    \[R_B^{new}=R_B+K\cdot(S_B-E_B)=2000+10\cdot(1-0.36)=2006\]

This result can be regarded as player B took 6 points away from player A as trophy.\\

Assuming another example, at this time we can set up an rarely seen scenario that the player A has a ranking score of 2100 and $K=10$ and player B has a ranking score of 1700 and $K=40$. Then $E_A=0.9$, $E_B=0.1$. Again let's assume player A loses, so $S_A=0$, $S_B=1$. The new ranking score will be:
\[R_A^{new}=R_A+K\cdot(S_A-E_A)=2100+10\cdot(0-0.9)=2091\]
\[R_B^{new}=R_B+K\cdot(S_B-E_B)=1700+40\cdot(1-0.1)=1736\]

This time, the reward for player B is several times greater than the loss of player A. Compared with competing against players of same level, being defeated by a statistically weaker player will cost more on the behalf of player A. This kind of scheme can respond quickly if the skill of one player does not match his rating score. When a player’s scores exceed (fall short of) their expected scores, the ELO rating system assumes that the player’s rating had been too low (high) at its outset, and so needs to be adjusted upwards (downwards). Ideally, the players under an ELO rating system will form a normal distribution\\

The above is the basic introduction of the ELO ranking system. Although ELO ranking system has already been improved and replaced by its variants, it is still playing an important role in the game ranking fields.\\

\section*{Problem}

When it comes to the context of nowadays' online competition games like Defense of the Ancients(Dota) or League of Legends(LOL), the competition is more complex and uncertain. On one hand, the match scheme becomes team versus team(usually 5 players for each team) rather than one player against another player. Therefore, how to formulate the performance of individual players into the competitiveness scocre of the whole team becomes the problem. On the other hand, another balance problem exists within each side of the team, which is how to ensure the equal contribution from each player to the winning. Assuming a 3v3 game, one team(A) is formed by players of competitiveness score {1200,400,400} and that of the other team(B) is {800,600,600}. Based on the total sum of the score, both of the two teams are expected to have the expectation of winning as 0.5. If team A wins, then the player 1000 is expected to have made the most of the contribution to the winning. This kind of contribution can be measured by KDA(kills/deaths/assists) in Dota or LOL.\\

\begin{figure}[h]
    \centering
    \includegraphics[width=0.4\textwidth]{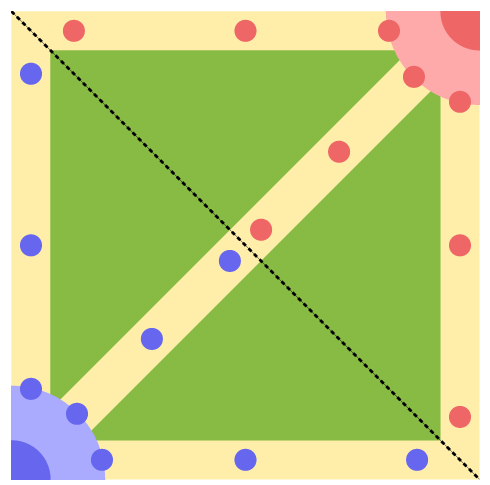}
    \caption{Example of the map in MOBA games}
    \label{map}
\end{figure}

Although winning the game is the goal of all the players, this kind of imbalanced gaming experience is not what gamers are expecting. In Multiplayer Online Battle Arena(MOBA) games, there are usually 3 lanes where players are playing 1v1 at the beginning of the game. See in Figure \ref{map}.

In the previous example, because of player 1000's overwhelming gaming skill, he should soon beat the player 600 he is facing and make his snowball to continuously upgrade his items. Then player 1000 is likely to take over the game until winning. Then if we dig into the feelings of each player:

\begin{enumerate}
    \item Player 1200 in A team: I did quite a good job. I smashed my opponent then led my team to the winning. However, my higher competitiveness score gives me a lower K value. My teammates are not making the same contribution as I did but they will gain more points than mine.
    \item Player 400 in A team: I was beaten by the opponent in my lane, but player 1000 is dominant enough to somehow "bring" me to the victory. I have little influence on the final result actually. Although the victory is encouraging, I am still expecting a more balanced competition.
    \item Player 800 in B team : The opponent clearly outperforms all the other players in this game, he should not be matched to this game. This game is unfair.
    \item Player 600 in B team who is facing player 400: Although I have beaten the enemy in my lane, I still lose the game because my teammate is too weaker compared with his competitor.
\end{enumerate}

In this example, contrast to the sum of competitive score, team A actually got more chance to win because the battle in MOBA is usually heavily influenced by the most dominant lane especially the mid and bottom lane\cite{10.1007/978-3-030-43722-0_27}. In other words, although the competitive scores are 1200,400,400, the importance of each player from team A in this game can be greatly redistributed to an extreme proportion like 1600,200,200. So how ELO has anything to do with the improper game matching?\\

Searching in Reddit using the keyword "ELO", we can find most of the posts are complaining about this kind of rating scheme. Although ELO system itself is not trying to introduce any unfairness to the game, it does to some extent suppress the player experience(PE). In MOBA ranking systems, players are rewarded with ranking points for being on the winning team rather than their individual performance. In the example above, although player 800 and 600 from B team are playing better than player 400s from A team, they still lose the game.\\

\begin{figure}[h]
    \centering
    \includegraphics[width=\textwidth]{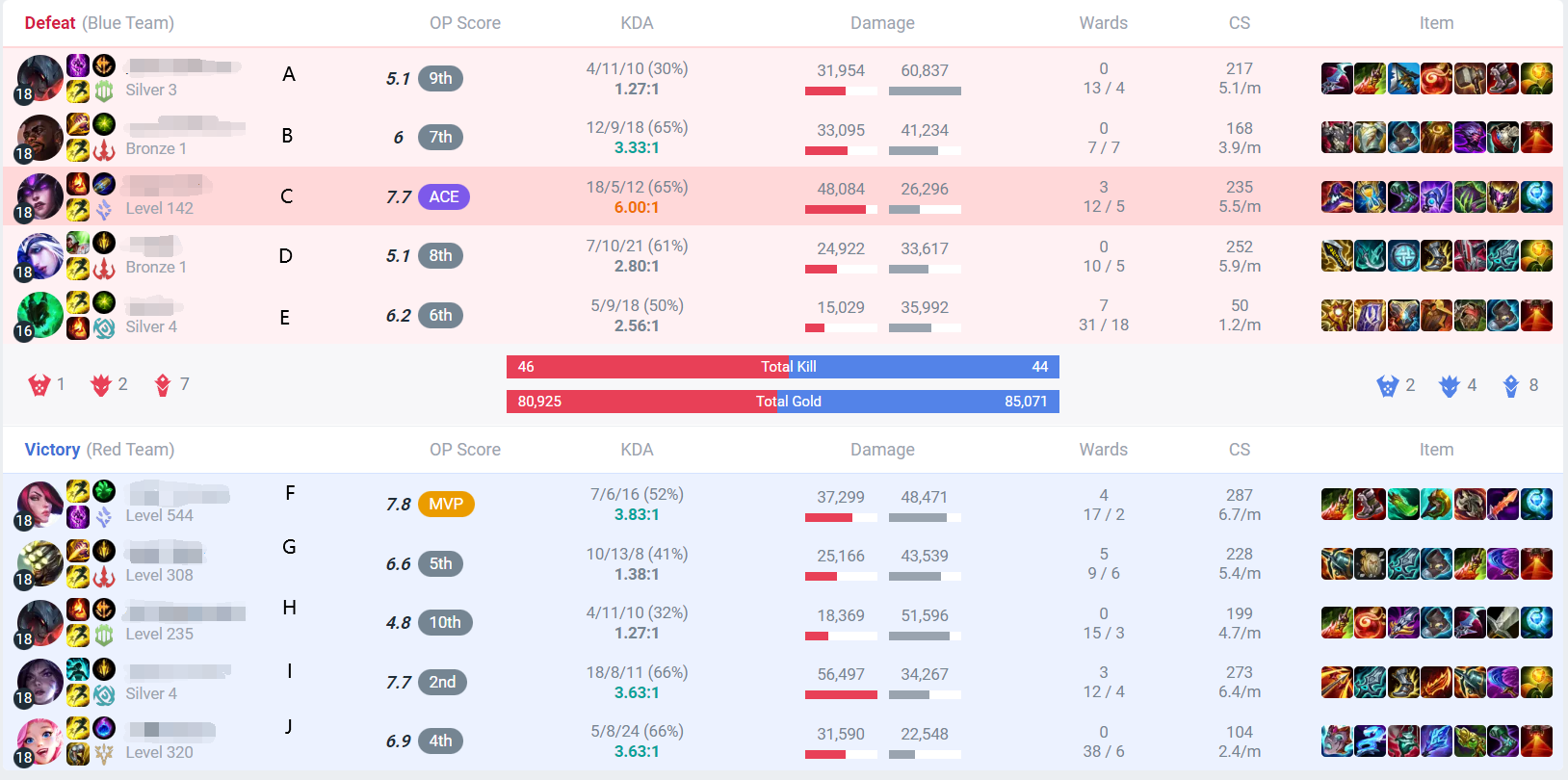}
    \caption{Example of A Real Match in LOL}
    \label{match}
\end{figure}

Taking a real game for example in Figure \ref{match}. This is the information from OP.GG\cite{opgg}, which is the most accessed platform for collecting and analyzing LOL players' data. Despite red team's victory in this match, player C in blue team has performed well enough. This kind of performance is indicated by his KDA and OP Score. The data of total kills and total gold also suggests that this game is numerically well-matched. Although player C has shown his skills during the game, the result turns out to be that he will lose his points while the relatively weaker players in red team will gain their ranking points. If there are hundreds of thousands, possibly millions of unskilled players, being placed on teams with hundreds of thousands of skilled players. 5v5 team games end up becoming lopsided affairs, as the unskilled players make vital mistakes that can cost the game. It’s not entirely unfair, as both teams in a 5v5 game will have a mixture of “casual” and “core” players in the same range of ranking points, but this creates matches that feel more like gambling, where the outcomes are determined more by luck than player skills. That is the most criticized mechanism of ELO rating in MOBA, because it is trying to make the team more ladder-shaped. A team will be made up with players of high performance and bad performance rather than players from a smaller range of true skill. In a word, ELO based scoring scheme is unfair to those players who are paying more efforts to the winning. If they lose, there will penalties regardless of how well they have performed during the game. If they win, they will be rewarded less points than their teammates because they have got a lower $K$ value.One thing worth noting is that the ELO rating score is different with ranking score in the MOBA games. The ELO rating score is derived from both player's ranking score and his recent performance and in game.

Why the MOBA game developers are still taking this scheme for game matching? Another rating and match scheme is called Match Making Rating(MMR). Under MMR scheme, a player will probably been matched with teammates and opponents who has approximately the same rating score. Then if your skill exceeds your current rating, you will gradually gain ranking points because your winning rate would be greater than 50\%, vice versa. In the end, you will stay at your ranking level stably with winning rate drifting slightly around 50\%. Therefore, the MMR is a smoother scheme than ELO. However, the developers would consider this scheme not encouraging enough to keep players' activity. Under MMR, a player would stay in his ranking level for quite long time and he would know his skill belongs to this level without potential chance of going upward or downward. This feeling of being bounded can kill the player's passion to continuously play ranking since he has already been aware that he will not make an encouraging progress neither a frustrating retrogress. But the situation is different if ELO rating system is implemented.\\

Under the circumstance that a player has a high ELO score, he would perform well in the game regardless of the final result. If he wins, he would get less rewards because his ELO score is assigned with a low $K$ value. Also, his ELO score would increase greatly so that he will be matched with more power teammates and opponents. In other worlds, the game will possiblely become hader for you. This will slow down the rate of progress for him to go upwards(taking more games and more time). If a player keeps playing well yet losing his game, his ELO score drops with his ranking score. Then he would be assigned with weaker opponents and winning the next series of games.\\ 

If a player does not play well recently, he would be assigned a low ELO score and be matched with teammates with high ELO score. If his team wins the game, he would be rewarded more points to encourage him keeping gaming because his $K$ value is high. If he loses, his elo score and ranking score would both drop and there will be weaker opponents in the next games to help him achieve victory.\\

In a word, the game developer's ELO system wants the winning rate of all the players to be only slightly higher or lower than 50\% regardless of the fact if this player's ranking level is located at his true skill level. Even if your skill far exceeds you ranking level, your winning rate will not be far above 50\%. Even if your skill falls below your ranking level, you will be assigned with powerful teammates so that it will be easier for your to win(you could also be the main blame for loss). That will make both going upward and downward slower. Drawing from this kind of scheme, ELO brings a paradox: the better you play, the harder for you to win.\\

\section*{Methods and Metrics}

\begin{figure}
    \centering
    \includegraphics[width=\textwidth]{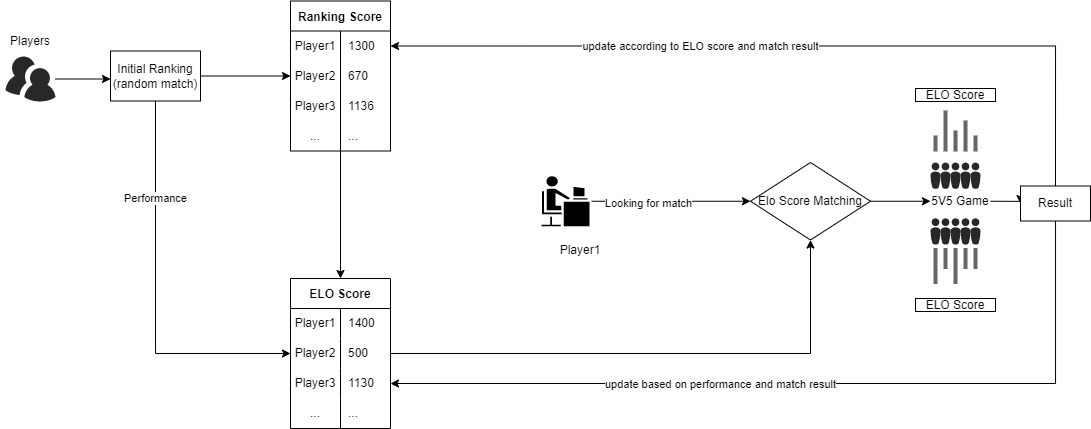}
    \caption{Matching System Workflow}
    \label{elo}
\end{figure}

A good rating system can reflect the player's true strength after a certain number of games. I am planning to run a simulation system of ELO rating. In the beginning, I will assume a group of players with random game strength from 100 to 2000 following normal distribution. After the players are assigned with their initial ranking score, they will then be matched with other players within a certain range of ELO rating score. We will see if the fitting curve of players' true skill and ranking level converges after a certain number of game rounds. The game result will be determined by the true competition power assigned to each player in the beginning of the simulation. See Figure \ref{elo} for the simulation matching system's workflow.\\

As what has been mentioned above, a dominant player in one lane will cast heavy influence on the result of the game so that his contribution to the team will rise beyond the proportion of gaming strength. To decide the result of a game(winning rate) in the simulation, I designed a dominance coefficient(DC) measured performance calculation method. The principle is to emphasize on the importance of one player's dominant skill based on inherent propriety of MOBA games. Equation \ref{performance} demonstrates how to calculate the individual performance score in a match, where $PS$ stands for performance score. $S$, $S_{oppo}$ equals to the player's game skill strength and that of his opponent at the same lane. 

\begin{equation}
    PS = S\cdot\frac{S}{S_{oppo}}
    \label{performance}
\end{equation}

For example, we have a game match-up like what is shown in Table \ref{dc}. The player strength will be multiplied with the delta coefficients, and the performance score will be used to calculate the winning possibility in this match. Based on ELO's prediction formulation, the winning rate for team A should be:

\[E_A=\frac{1}{1+10^{(1527.2-1617.4)/400}}\approx0.627\]

\begin{center}
\begin{table}
\begin{tabular}{|c|c|c|c||c|c|c|}
\hline
    \multicolumn{7}{|c|}{Team A vs Team B} \\
    \hline
     Lane & Strength A &  DC & Performance A & Strength B &  DC & Performance B\\
     \hline
     Top & 1600 & 1.143 & 1829 & 1400 & 0.875 & 1225\\
     Juggle & 1750 & 1.167 & 2041 & 1500 & 0.857 & 1286\\
     Mid & 1000 & 0.5 & 500 & 2000 & 2.000 & 4000\\
     Bottom & 1300 & 1.3 & 1690 & 1000 & 0.769 & 769\\
     Support & 1300 & 1.3 & 1690 & 1000 & 0.769 & 769\\
     Average & 1380 &  & 1572.8 & 1380 &  & 1609.8\\
    
\hline
\end{tabular}
\caption{Example of DC-Devised Score}
\label{dc}
\end{table}
\end{center}

Winning is always expected, while the player's experience(PE) is important as well. To find whether ELO is indeed detrimental to players' experience, we need to find some quantities support. Iida et al. proposed a Motion in Mind model\cite{motioninmind} derived from Game Refinement Theory\cite{10.1007/978-3-319-08189-2_22}. According to Iida's research, the attractiveness of a game can be measured similarly by real-world laws of physic. Table \ref{motion} shows the correspondence between game refinement theory and physics. \\

\begin{table}
\begin{center}
\begin{tabular}{|c|c||c|c|}
\hline
     Notation & Game & Notation & Physics \\
     \hline
     $y$ & solved uncertainty & $x$ & displacement\\
     $t$ & total score or game length & $t$ & time\\
     $v$ & winning rate(v=p) & $v$ & velocity\\
     $m$ & winning hardness($m=1-p$) & $M$ & mass\\
     $a$ & acceleration in mind & $g$ & gravitational acceleration\\
     $\overrightarrow{p}$ & momentum of game & $\overrightarrow{p}$ & momentum\\
     $E_p$ & potential energy of game & $U$ & potential energy\\
\hline
\end{tabular}
\caption{Analogical link between game refinement theory and physics.}
\label{motion}
\end{center}
\end{table}

In another work from Thavamuni et al.\cite{THAVAMUNI2023100523}, they adapted Motion in Mind model to MOBA games. Following Iida and Thavamuni's idea, I made a simplified assumption that a player's experience depends on how well he has performed comparing with his teammates and the final result. The equation to calculate $v$ is described in Equation \ref{exp}. According to this velocity value, we can get other informative data to analyze the player's experience under different rating scheme like mass in mind, energy in game based on Iida's previous work\cite{motioninmind} .


\begin{equation}
     v =\begin{cases}
        \frac{PS}{PS_T} \cdot \frac{\text{Points Earned}}{\text{Average Points Earned}} & \text{if $win$}\\
        0 & \text{if $loss$}
   \end{cases}
   \label{exp}
\end{equation}\\


Finally, I would test a more proper reward system in the simulation. My proposed scheme follows the principle that an effortful player will be awarded more ranking credits based on his contribution to the winning. Vice versa, the effortful player will be cost less points if he loses his game. I will compare the converge condition of the proposed scheme with that of ELO scheme. Along with players' experience will be compared to check if my proposed rating method outperforms the ELO method.

\section*{Experiment and Results}

\subsection*{Initialization}
In the simulation experiment, there are two ranking tables for players. One table shows the apparent ranking score, which in real MOBA games is recorded as ladder score for everyone to see. Another is ELO ranking table, which records the performance of a player and is hidden from all players. In the beginning, 2000 virtual players are generated and assigned with a game skill strength from 210 to 2000. The initial ladder score and ELO score are calculated as Equation \ref{ini} and \ref{elo_e}, where $\Delta$ is a random integer from -500 to 500:

\begin{equation}
    Ladder = Strength + \Delta
    \label{ini}
\end{equation}
\begin{equation}
    ELO = Strength + \Delta
    \label{elo_e}
\end{equation}

\begin{figure}
    \centering
    \includegraphics[width=\textwidth]{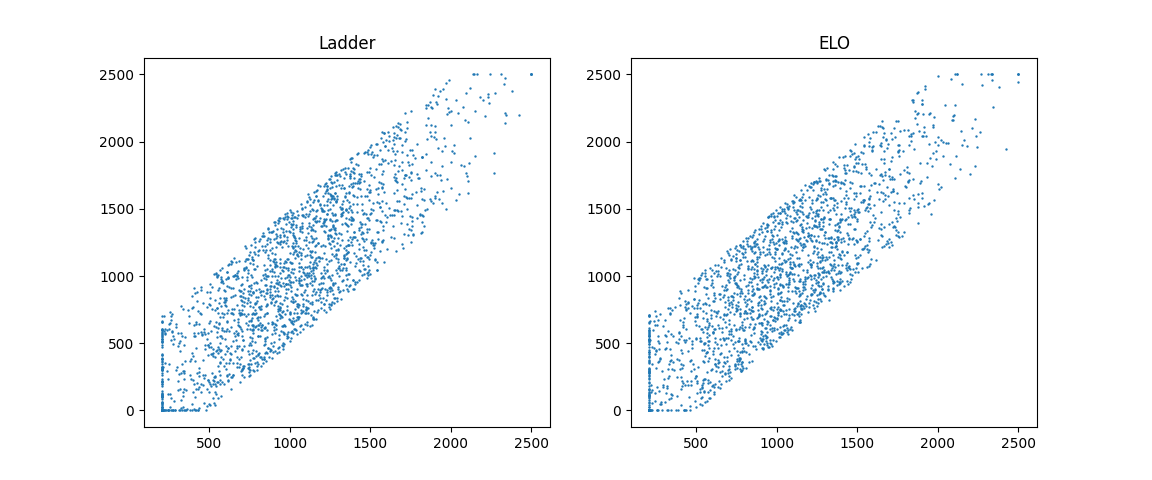}
    \caption{Initial Ladder and ELO Score}
    \label{initial}
\end{figure}

The initial distribution of ladder score and ELO score are shown in Figure \ref{initial}.

\subsection*{Matching Strategy}

The system will randomly pick two players as the core players of team A and team B, and the ELO scores of these two players should be as close as possible. At first, the difference between the two players' ELO scores is set as 0, but the threshold will be eased if the match keeps failing (measured by loop counts). Then, following the same pattern, the system will build team A and team B.

\subsection*{Scoring Policy}

The ladder score will be updated according to the game result and the player's $K$ value, where $K$ value is drawn from the player's ELO score according to Equation \ref{kvalue}. The updating method is same as the design in original ELO rating scheme. $S_A$ and $E_A$ represent for the game result and winning expectancy in Equation \ref{ladderupdate}:

\begin{equation}
    K=35-5\cdot\lfloor\frac{ELO}{400}\rfloor
    \label{kvalue}
\end{equation}
    
\begin{equation}
    Ladder_{new} = Ladder_{old} + K\cdot(S_A-E_A)
    \label{ladderupdate}
\end{equation}
\\
The ELO score will be updated based on the game result and the player's performance in the game. Equation \ref{eloupdate} shows different updating strategies for victory and defeat, where $PS$ is each player's performance score and $PS_T$ represents for the team's average performance score. This pattern can increase "good" players' ELO score rapidly if they keep winning, and it will also suppress the decreasing of ELO score if this "good" player is keeping losing his game.

\begin{equation}
ELO_{new} =\begin{cases}
        ELO_{old}+\frac{PS}{PS_T} \cdot K\cdot(S_A-E_A) & \text{if $win$}\\
        ELO_{old}+\frac{PS_T}{PS} \cdot K\cdot(S_A-E_A) & \text{if $loss$}
        \end{cases}
    \label{eloupdate}
\end{equation}

\subsection*{Simulation Results}

\begin{figure}[H]
    \centering
    \includegraphics[width=0.85\textwidth]{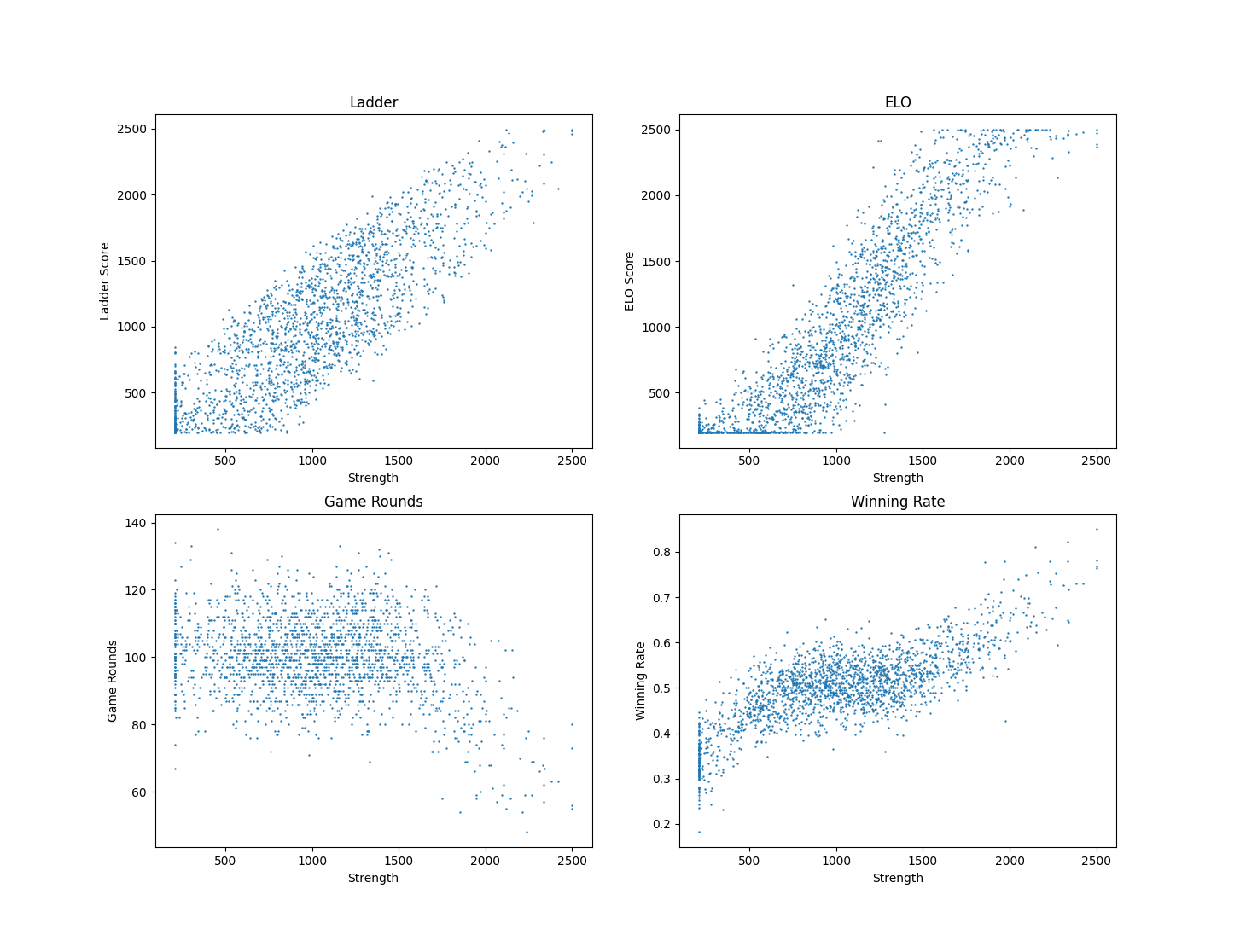}
    \caption{After 10000 Matches}
    \label{fig:10000}
\end{figure}

After 10000 games conducted, the converge condition is shown in Figure \ref{fig:10000}.\\

\begin{figure}[H]
    \centering
    \includegraphics[width=0.85\textwidth]{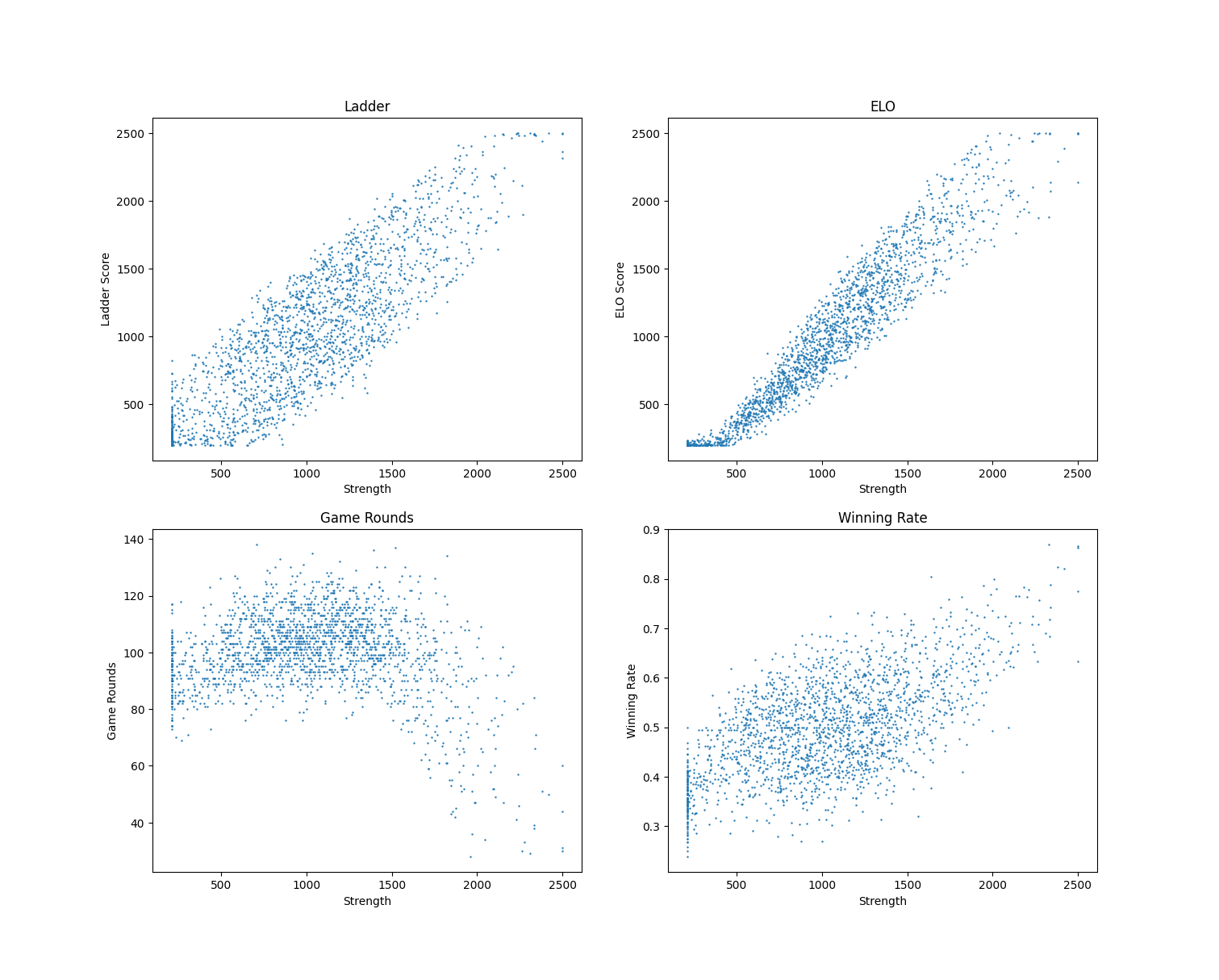}
    \caption{After 20000 Matches}
    \label{fig:20000}
\end{figure}

After 20000 games conducted, the converge condition is shown in Figure \ref{fig:20000}.\\

\begin{figure}[H]
    \centering
    \includegraphics[width=0.85\textwidth]{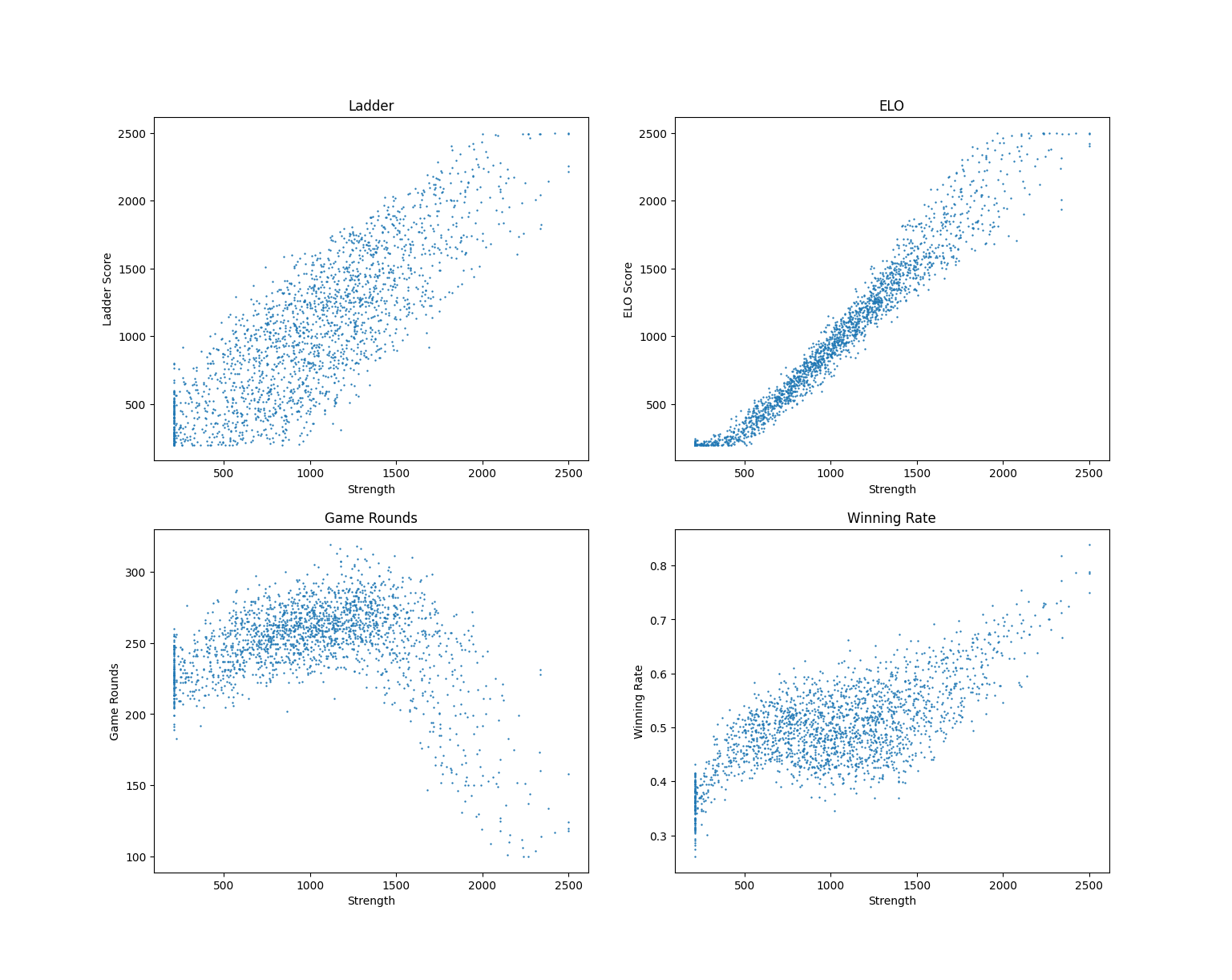}
    \caption{After 50000 Matches}
    \label{fig:50000}
\end{figure}

After 50000 games conducted, the converge condition is shown in Figure \ref{fig:50000}.\\

\begin{figure}[H]
    \centering
    \includegraphics[width=0.85\textwidth]{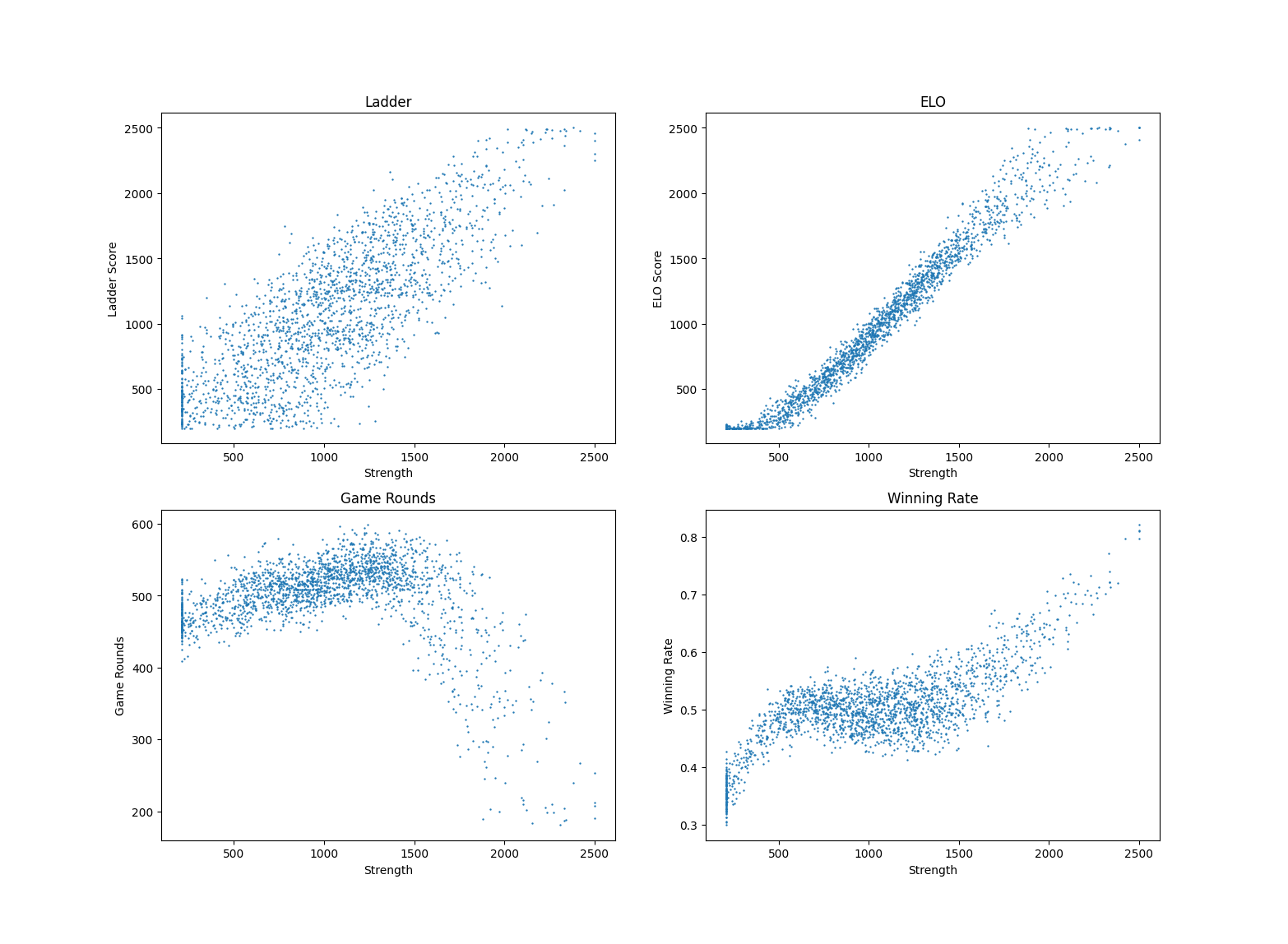}
    \caption{After 100000 Matches}
    \label{fig:100000}
\end{figure}

After 100000 games conducted, the converge condition is shown in Figure \ref{fig:100000}.\\

From the figures, we can clearly find that ELO score converges much faster than the ladder. In other words, ELO score actually better reflects the player's strength. That justifies the previous assumption that utilizing ELO scheme will motivate players to spend more time on the game because uncertainty is introduced into the matching system so that the player will not smoothly move to his position in the ladder system.

\subsection*{Proposed Matching Model}

In my proposed rating system, the ranking score will be updated based on both game result and player's performance. This pattern is similar to ELO's updating method, but without $K$ value assigned. A player with higher performance score will be rewarded more ladder points comparing with his less competitive teammates, and it will cost him less penalty if he plays well but still loses the game. Equation \ref{eq:p} shows how to update the ladder score in the proposed scheme.

\begin{equation}
    Ladder_{new} =\begin{cases}
        Ladder_{old}+\frac{PS}{PS_T} \cdot 20 & \text{if $win$}\\
        Ladder_{old}-\frac{PS_T}{PS} \cdot 20 & \text{if $loss$}
        \end{cases}
        \label{eq:p}
\end{equation}\\

\begin{figure}[H]
    \centering
    \includegraphics[width=\textwidth]{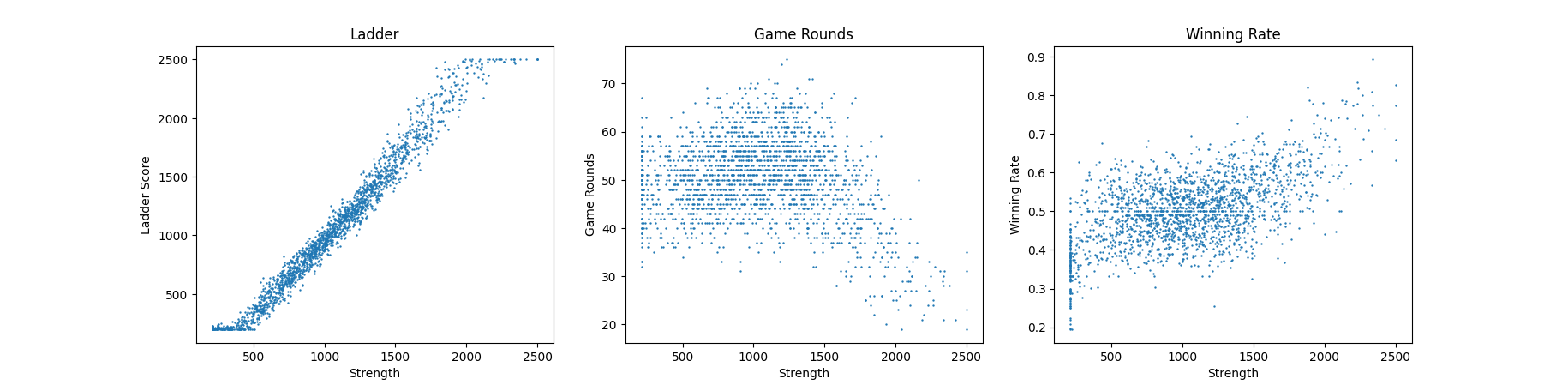}
    \caption{After 10000 Matches}
    \label{fig:p_10000}
\end{figure}

\begin{figure}[H]
    \centering
    \includegraphics[width=\textwidth]{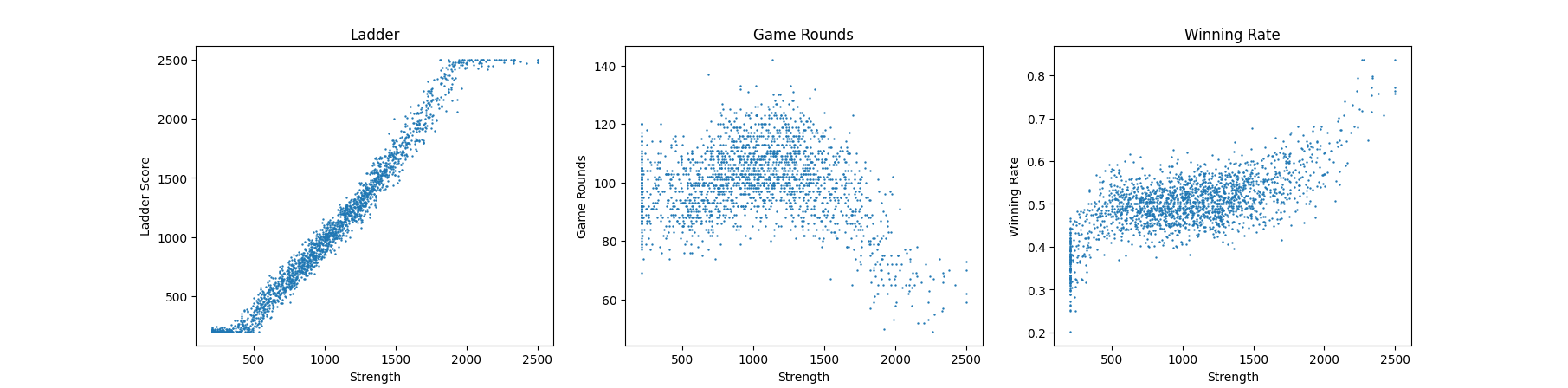}
    \caption{After 20000 Matches}
    \label{fig:p_20000}
\end{figure}

The matching strategy follows the same design as the previous simulation. Figure \ref{fig:p_10000} and \ref{fig:p_20000} is the simulation results after 10000 and 20000 rounds of matches.

It can be drawn from the simulation result that the proposed matching and ranking scheme converges much faster than the ELO scheme. The players' ladder score matches the position of his true game strength only after 10000 total games. Therefore, this ranking scheme is more precise and efficient than the ELO scheme.

\subsection*{Experience Analysis}

\begin{figure}[H]
    \centering
    \includegraphics[width=0.85\textwidth]{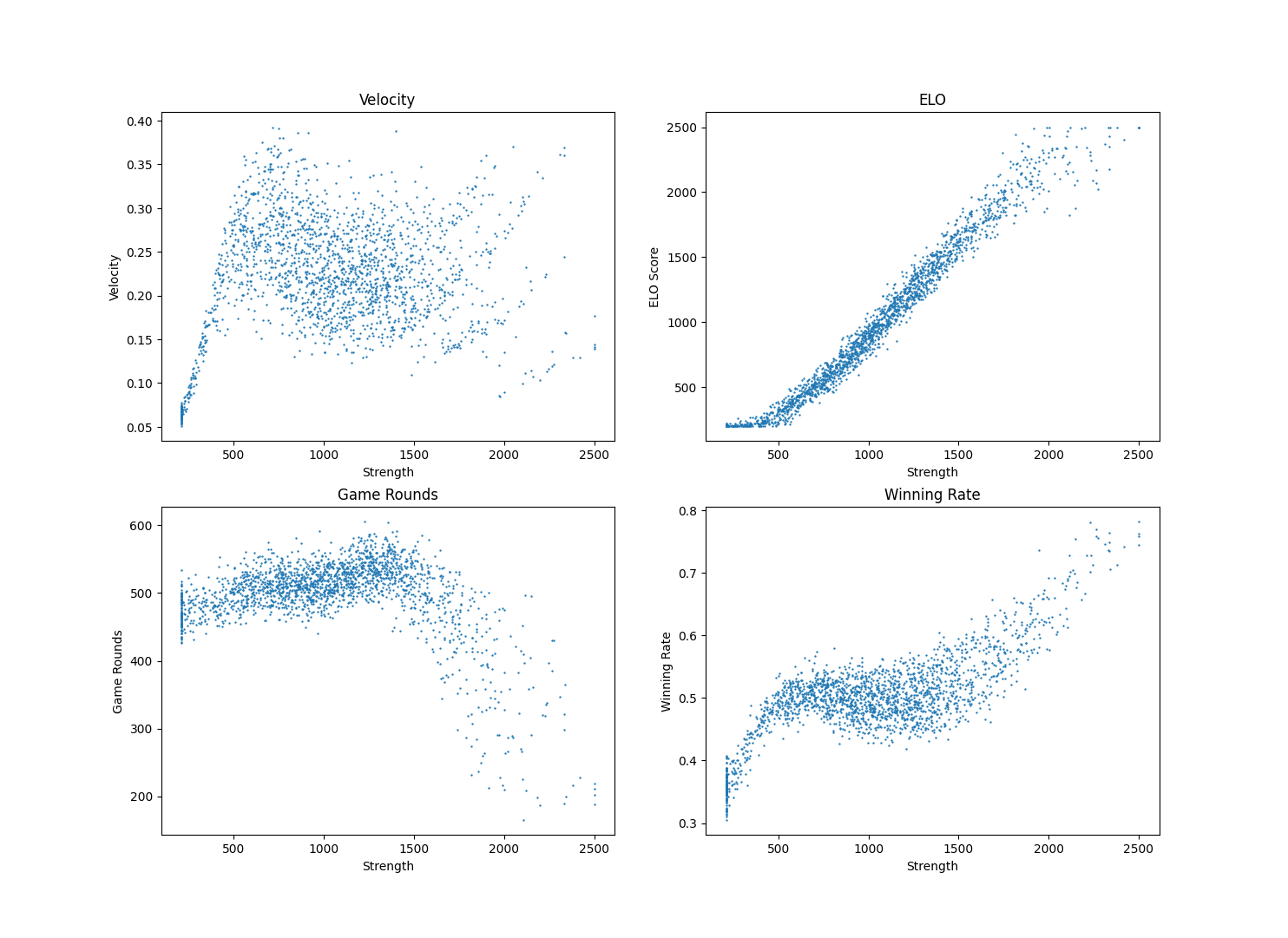}
    \caption{Experience under ELO Matching Model}
    \label{v_elo}
\end{figure}

\begin{figure}[H]
    \centering
    \includegraphics[width=0.85\textwidth]{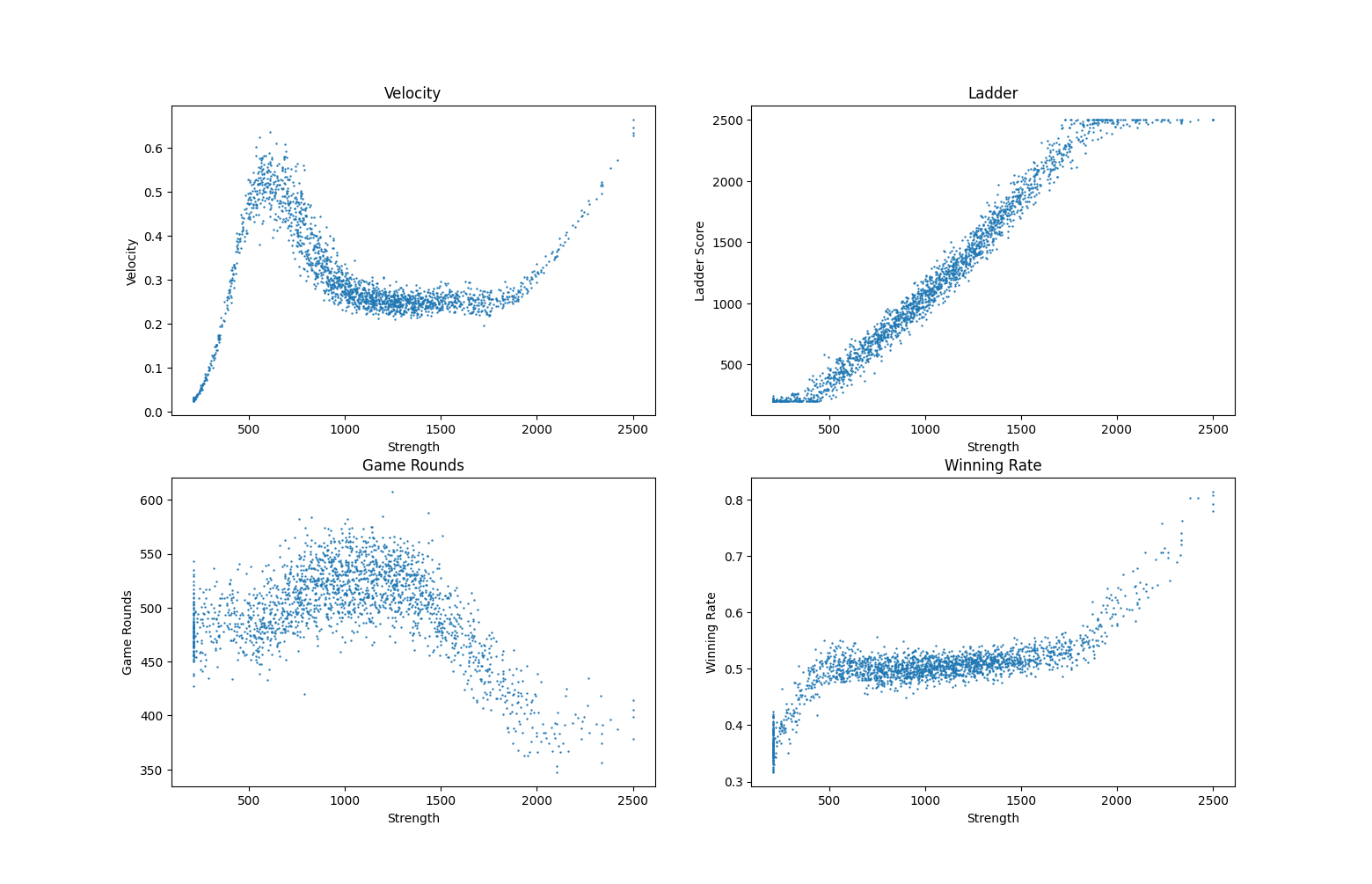}
    \caption{Experience under Proposed Matching Model}
    \label{v_p}
\end{figure}

For evaluating players' experience under different rating scheme, according to Equation \ref{exp}, the velocity under ELO matching scheme and the proposed matching scheme is shown in Table \ref{tb:exp}. And the velocity distribution is shown in Figure \ref{v_elo} and \ref{v_p}.

\begin{table}
\begin{center}
\begin{tabular}{|c|c|c|c|}
\hline
     Matching Mode & Mass & Velocity & Potential Energy\\
     \hline
     ELO & 0.558 & 0.442 &  0.218\\
     Proposed & 0.404 & 0.596 & 0.287\\ 
     \hline
\end{tabular}
\caption{Comparison of Motion in Mind Values}
\label{tb:exp}
\end{center}
\end{table}

According to the result, the proposed matching scheme still outperforms ELO scheme in the context of player's experience.

\section*{Conclusion and Future Work}

ELO rating scheme aims to minimizing the gap between the player's true skill and his ranking score like all the other rating schemes. Besides, the original design of ELO is fair and efficient. The utilization of strength based scoring policy and $K$ value accelerates the convergence of player's true game strength and his ranking level. What is disliked by the MOBA game players is actually performance based ELO. Under that kind of rating scheme, the difficulty of game will go ahead of the player's ranking score since ELO will amplify the influence player's of performance while building a match. A player will encounter a much more challenging match if he plays well in the previous game, vice versa. Although this scheme is not helping with the players' experience, it slows the convergence of players' ranking score. To some extent, ELO scheme brings controllable uncertainty into the matching strategy so that the players will not feel bored after they reach their true position in the ranking system. On behalf of effort and reward, ELO is not a proper rating scheme because performance and gained ranking points are not positively correlated. Compared with ELO, the proposed rating scheme is more reasonable and efficient. On the other hand, ELO rating scheme is able to keep the uncertainty of matches after a certain extent of convergence. The future work can revolve around finding that turning point. Namely, the proposed rating scheme can be implemented at the beginning of a ranking season. When players' strength and ranking score reach a certain level of convergence, the matching system can turn into ELO model for motivating players to keep playing ranking. Therefore, the future work is planed use the Motion in Mind theory to model the player's experience on the aspect of both reward and attractiveness so that we can find the proper time point to switch from proposed rating scheme to ELO based rating scheme.

\bibliographystyle{unsrt}
\bibliography{refs}

\begin{thebibliography}{1}

\bibitem{elo1978rating}
Arpad~E. Elo.
\newblock {\em The Rating of Chessplayers, Past and Present}.
\newblock Arco Pub., New York, 1978.

\bibitem{enwiki:1050416173}
{Wikipedia contributors}.
\newblock Kenneth harkness --- {Wikipedia}{,} the free encyclopedia, 2021.
\newblock [Online; accessed 5-December-2022].

\bibitem{10.1007/978-3-030-43722-0_27}
Alberto~Mateos Rama, Victor Rodriguez-Fernandez, and David Camacho.
\newblock Finding behavioural patterns among league of legends players through
  hidden markov models.
\newblock In Pedro~A. Castillo, Juan~Luis Jim{\'e}nez~Laredo, and Francisco
  Fern{\'a}ndez~de Vega, editors, {\em Applications of Evolutionary
  Computation}, pages 419--430, Cham, 2020. Springer International Publishing.

\bibitem{opgg}
Bughunter-summoner stats-league of legends.
\newblock
  \url{https://www.op.gg/summoners/jp/BugHunter/matches/CahDN-kFxcGFQ4mn4ydnNWZqhQ9DSZTO/1669121940000}.
\newblock Accessed: 2022-12-7.

\bibitem{motioninmind}
Hiroyuki Iida and mohd nor~akmal Khalid.
\newblock Using games to study law of motions in mind.
\newblock {\em IEEE Access}, 8:1--1, 01 2020.

\bibitem{10.1007/978-3-319-08189-2_22}
Arie~Pratama Sutiono, Ayu Purwarianti, and Hiroyuki Iida.
\newblock A mathematical model of game refinement.
\newblock In Dennis Reidsma, Insook Choi, and Robin Bargar, editors, {\em
  Intelligent Technologies for Interactive Entertainment}, pages 148--151,
  Cham, 2014. Springer International Publishing.

\bibitem{THAVAMUNI2023100523}
Sagguneswaraan Thavamuni, Mohd Nor~Akmal Khalid, and Hiroyuki Iida.
\newblock What makes an ideal team? analysis of popular multiplayer online
  battle arena (moba) games.
\newblock {\em Entertainment Computing}, 44:100523, 2023.

\end{thebibliography}

\end{document}